\documentstyle[12pt,epsfig]{article}
\newcommand{\m}{\medbreak}

\newcommand{\no}{\noindent}
\newcommand{\EQ}{\begin{equation}}
\newcommand{\eq}{\end{equation}}
\newcommand{\EQA}{\begin{eqnarray}}
\newcommand{\eqa}{\end{eqnarray}}

\newcommand{\AR}{\renewcommand {\arraystretch}{1.5}
\begin{array}{l}}
\newcommand{\bAR}{\renewcommand {\arraystretch}{2}
\begin{array}{l}}
\newcommand{\ARc}{\renewcommand {\arraystretch}{1.5}
\begin{array}{c}}
\newcommand{\bARc}{\renewcommand {\arraystretch}{2}
\begin{array}{c}}
\newcommand{\ar}{\end{array} \renewcommand {\arraystretch}{1}}

\newcommand{\ee}{\mbox{$e^+e^-\ $}}

\newcommand{\ET}{\mbox{$E_T\ $}}

\newcommand{\AL}{\mbox{$A_{L}\ $}}

\newcommand{\r}{\rightarrow}

\newcommand{\Z}{$Z^{\circ}\ $}
\newcommand{\ZP}{$Z'\ $}
\newcommand{\WP}{$W'\ $}

\begin{document}
\begin{titlepage}
\vspace{0.2in}
\vspace*{1.5cm}
\begin{center}
{\large \bf Discovery potential for New Physics \\in view of the
RHIC-Spin upgrade
\\} 
\vspace*{0.8cm}
{\bf P. Taxil} and {\bf J.-M. Virey}  \\ \vspace*{1cm}
Centre de Physique Th\'eorique$^{\ast}$, C.N.R.S. - Luminy,
Case 907\\
F-13288 Marseille Cedex 9, France\\ \vspace*{0.2cm}
and \\ \vspace*{0.2cm}
Universit\'e de Provence, Marseille, France\\
\vspace*{1.8cm}
{\bf Abstract} \\
\end{center}
In view of a possible upgrade of the RHIC-Spin program
at BNL, concerning both the machine and the detectors,
we give some predictions concerning the potentialities
of New Physics detection with polarized proton beams.
We focus on parity-violating asymmetries in one-jet
production due to contact terms or to a new leptophobic
neutral gauge boson. We comment on the main 
uncertainties and we compare with unpolarized searches
at Tevatron.
 \\

\vfill
\begin{flushleft}
\m\no
Number of figures : 1\\

\m\no
September 2001\\
CPT-01/P.4224\\
\m\no
anonymous ftp or gopher : cpt.univ-mrs.fr

------------------------------------\\
$^{\ast}$Unit\'e Propre de Recherche 7061
 \\
E-mail : taxil@cpt.univ-mrs.fr ; virey@cpt.univ-mrs.fr
\end{flushleft}
\end{titlepage}

\section{Introduction}
\indent
\m
There is a growing interest on the physics program which will be 
achieved at RHIC-Spin, that is at the 
Brookhaven National Laboratory Relativistic Heavy Ion Collider (RHIC), 
running in the  polarized $\vec p \vec p$ mode.

Actually, during the year 2001 the RHIC-Spin Collaboration (RSC) will 
perform the first polarized
run at a c.m. energy of $\sqrt s = 200 $ GeV and
a luminosity of a few $10^{30} cm^{-2}s^{-1}$.


The nominal energy of $\sqrt s = 500 $ GeV and luminosity
${\cal L}\ =\, 2. 10^{32} cm^{-2}s^{-1}$ should be reached
in the early months of 2003, allowing an exposure
of 800 $pb^{-1}$ in four months of running.

Physics at RHIC-Spin has been extensively covered in a recent review
paper \cite{Bunce}, where many references can also be found
(see also \cite{Spin2000} ).
The first part of the program will include precise measurements
of the polarization of the gluons, quarks and sea-antiquarks
in a polarized proton. This will be done thanks to well-known
Standard Model processes : direct photon, $W$ and $Z$ production, 
Drell-Yan pair production, heavy-flavor production
and the production of jets.
The helicity structure of perturbative QCD will be thoroughly
tested at the same time with the help of Parity Conserving
(PC) double spin asymmetries.

It has been first noticed more than ten years ago \cite{Tannenbaum} that
the production of high $E_T$ jets from polarized protons
could allow to pin down
a possible new interaction, provided that parity is 
violated in the subprocess. 

Since QCD is parity conserving and dominates the process,
according to the Standard Model (SM),
the expected Parity Violating (PV) spin asymmetry 
in jet production should come from tiny electroweak effects.
Hence, a net deviation from the small expected Standard Model
asymmetry could be a clear signature of the presence of
New Physics.
 
Due to the energy reach of the machine
the New Physics scale should not be too high 
to yield a contribution : fortunately 
some scenarios are still allowed by present data, in particular
the existence of a new weak force belonging uniquely to
the quark sector.

In previous papers, we have explored
the very phenomenological case of a PV contact
interaction between quarks \cite{TVCT}, various situations
with a new \ZP with nearly zero couplings
to leptons (the so-called leptophobic $Z'$) \cite{TVZprime}
and also a scenario with a right-handed \WP 
decaying into quarks in the case of a very massive right-handed
neutrino \cite{TVWprime}.

In this letter we will explore the potentialities of RHIC-Spin in view
of the two kinds of possible upgrades \cite{Saito}. The improved
machine could reach
$\sqrt s = 650 $ GeV with an integrated luminosity
$L = 20 fb^{-1}$ in a few months running and the STAR detector could
greatly improve the angular coverage with new end-caps. 
We compare also with the limits which could
be obtained with the (unpolarized) Tevatron in Run-II.
Concerning theoretical uncertainties, we comment the situation
on higher-order calculations when they are available.

\m

\section{Sources of PV effects in jet production}

The production of high $E_T$ jets is dominated by QCD, in particular
by quark-quark scattering. The existence of $W$ and \Z adds a small standard
contribution to the cross section \cite{AbudBaurGloverMartin}.
 On the other hand
the interference of weak amplitudes with QCD amplitudes will
be the main Standard source of PV helicity asymmetries in this process.
A peak in the asymmetries at $E_T \approx M_{W,Z}/2$ is also the main signature
for a pure electroweak contribution.

All the tree-level polarized cross sections for these standard
subprocesses can be found in
Ref. \cite{BouGuiSof}. Predictions using updated polarized partonic distributions
can be found in \cite{TVZprime} or \cite{Bunce}.
\\
The effects of some possible Non Standard PV interactions have been studied in 
recent years :

- First \cite{TVCT} one can think to a simple phenomenological contact interaction
which could represent the consequences of quark compositeness. Such 
(color singlet and isoscalar) terms are
usually parametrized following Eichten et al. \cite{EichtenEHLQ} :

\EQ\label{Lcontact}
{\cal L}_{qqqq} = \epsilon \, {g^2\over {8 \Lambda^2}} 
\, \bar \Psi \gamma_\mu (1 - \eta \gamma_5) \Psi . \bar \Psi
\gamma^\mu (1 - \eta \gamma_5) \Psi
\eq
\noindent
where $\Psi$ is a quark doublet, $\epsilon$ is a sign and $\eta$ 
can take the values $\pm 1$ or 0. $g$ is a new strong coupling
constant usually normalized to $g^2 = 4\pi\,$ and $\Lambda$ is
the compositeness scale.

In the following we will consider the $LL^-$ case with Left-handed
chiralities ($\eta = 1$) and constructive interference with
QCD amplitudes which corresponds to $\epsilon = -1$.

- Second, we can consider some new neutral gauge bosons
with general Left and Right-handed couplings 
to each given quark flavor $q$:
\EQ
\label{lag}
{\cal L}_{Z'} = \kappa {g\over 2 \cos \theta_W} Z'^{\mu}{\bar q} \gamma_\mu[ C^q_{L}
(1 - \gamma_5) \; +\; C^q_{R} (1 + \gamma_5) ] q
\eq \no
the parameter $\kappa = g_{Z'}/g_Z$ being
of order one. 
For a recent review on \ZP phenomenology (in the context of \ee collisions), one can consult 
\cite{Leike}. 
A particular class
of models, called leptophobic \ZP, is poorly constrained
by present data since they evade the LEP constraints. 
Such models appear in several string-inspired scenarios
\cite{LopezNanopoulos,LykkenBabu}. Non supersymmetric models can also
be constructed \cite{NoSusy}. Other references can be found in  \cite{TVZprime}.
In addition, it was advocated in \cite{Cvetic} that such
a boson could appear with a mass close to the electroweak scale 
and a mixing angle to the
standard \Z close to zero.

In this letter we will focus for illustration on the
 flipped-SU(5) model of Lopez and Nanopoulos 
\cite{LopezNanopoulos}  (model A of \cite{TVZprime}) in which parity is maximally
violated in the up-quark sector. 
Therefore the couplings in eq.(2) take the following values :
$C_L^u = C_L^d = - C_R^d = 1/(2{\sqrt 3})$ and $C_R^u = 0$, the
ratio $\kappa$ being a free parameter.
In this scenario,  
95\% of the new PV effect will come from
the interference between the \ZP exchange and the one-gluon exchange amplitudes
in the scattering of $u$ quarks in the $t$-channel.\\

\m
\section{Results}
\indent
\m
For Spin experiments, the most important quantities in practice are not
the polarized cross sections themselves, but the spin asymmetries.

At RHIC, running in the $\vec p \vec p$ mode, it will be possible to measure 
with a
great precision  the single PV asymmetry \AL :
\EQ
A_L \; =\; {d\sigma_{(-)}-d\sigma_{(+)}\over 
d\sigma_{(-)}+d\sigma_{(+)}}
\eq
\no where only one of the proton is polarized, 
or
the double helicity PV asymmetry :
\EQ
\label{ALLPVdef}
A_{LL}^{PV} ={d\sigma_{(-)(-)}-d\sigma_{(+)(+)}\over 
d\sigma_{(-)(-)}+d\sigma_{(+)(+)}}
\eq
\noindent where both polarizations are available.
In the above quantities the signs  $\pm$ refer to the helicities of the colliding
protons. The cross section $d\sigma_{(\lambda_1)(\lambda_2)}$ means the one-jet
production cross section in a given helicity configuration, 
$p_1^{(\lambda_1)}p_2^{(\lambda_2)} \r jet + X$, estimated at  
some $\sqrt{s}$
for a given jet transverse energy \ET, integrated over a pseudorapidity interval 
$\Delta \eta \,$ centered at $\eta\,=\,0$.  
In fact, both quantities will exactly yield the same amount of information.
From now we will discuss only the single PV asymmetry.
\m
All the
present calculations use polarized parton distribution functions
$\Delta f_i(x,Q^2)$'s
which have been parametrized from deep-inelastic data
e.g. GRSV distributions \cite{GRSV}. The polarized quark distributions
$\Delta q(x,Q^2)$
which play a dominant role in our calculation at high $E_T$
are the most reliable : in any case they will be much better
measured soon 
thanks to the first part of the RHIC-Spin program itself.
\m
We give in Table 1 the 95 \% C.L. limits on $\Lambda \equiv \Lambda_{LL^-}$
(eq.1) one gets, at lowest order, from a comparison between the 
SM asymmetry $A_L$ and the Non-Standard one.
 We have taken into account the
statistical error, which for small asymmetries is given by :
\EQ
\Delta A_L = {1 \over P}\,{1 \over {\sqrt N}}
\eq
\no
where $P$ is the degree of polarization of the beams, expected to be $P = 0.7$.
Systematics are assumed to be low \cite{Bunce} (see comments below), and we
have taken the conservative value
$\delta_{syst} \equiv (\Delta A)_{syst}/A = 10\%$.

One can compare the bounds at $\sqrt s = 500$ GeV and $L = 0.8 fb^{-1}$ with
the ones after the energy and/or luminosity upgrade. $4 fb^{-1}$ 
($100 fb^{-1}$) represents
5$\times\,$4 months running with the presently designed (future) nominal luminosity.
On the other hand, the cross section being essentially flat 
in rapidity in the interval which is accessible to experiment, an increase
in rapidity from $\Delta\eta =1$ to 2.6 is equivalent to a substantial
increase in luminosity.\\

 Since the statistical error goes like (1/$P$)$(1/\sqrt N)$, reducing the
degree of polarization $P$ by a factor $\epsilon$ is equivalent
to a factor $\epsilon^2$ in luminosity. In practice, varying the designed value
$P=0.7$ by 10\% will change the limits by roughly 6\%.

\begin{center}
\begin{tabular}{|c|c||||c||c| }
\hline
$\sqrt{s}$ ($GeV$)& L ($fb^{-1}$) & $\Lambda (\Delta \eta = 1)$
& $\Lambda  (\Delta \eta = 2.6 )$\\
\hline
\hline
\hline
500 & 0.8 & 3.2 & 4.0\\
\hline
\hline
500 & 4 & 4.55 & 5.5\\
\hline
500 & 20 & 6.15 & 7.0\\
\hline
500 & 100 & 7.55 & 8.30\\
\hline
650 & 4 & 5.2 & 6.3 \\
\hline
650 & 20 & 7.05 & 8.10 \\
\hline 
650 & 100 & 8.75 & 9.5 \\
\hline
\end{tabular} 
\end{center}
\begin{center}
Table 1 : Limits on $\Lambda_{LL^-}$, in TeV,
at 95\% CL with $\delta_{syst} = 10\% $, $P=0.7$.
\end{center}

This table can be compared with the last published analysis
  of the D0 experiments at Tevatron \cite{D0lambda}: 
$\Lambda > 2.2$ TeV (95\% C.L.) from the dijet mass cross section.
From these figures we have extrapolated
a limit at Tevatron of 3.2 TeV (3.7 TeV) with a 1 $fb^{-1}$ (10 $fb^{-1}$) exposure.
\m
Turning now to the case of a leptophobic \ZP, we present in
Fig.1 the constraints on the parameter space ($\kappa, M_{Z'}$) obtained
from $A_L$ in the flipped SU(5) model. 
The dotted curves correspond to ${\sqrt s} = 500$ GeV and the
dashed curves to ${\sqrt s} = 650$ GeV. From bottom to top
they correspond to an integrated luminosity
$L = 1,10,100 fb^{-1}$. It appears that the increase in luminosity
is more efficient than the increase in energy. 
Therefore the high
luminosity scenario has to be supported even if the RHIC $pp$
c.m. energy remains at its "low" value.

We display also in Fig.1 the inferred constraints coming 
from the published results of UA2 \cite{UA2}, CDF \cite{CDFjets2}
and D0 \cite{D0jets} experiments. The form of the forbidden areas
result from a combination of statistical and systematic errors.
For high $M_{Z'}$ one looks for some unexpected high-$E_T$
jet events and the main uncertainty is statistical in nature.
For instance, the upper part of the "CDF area" is well below
the one of D0 because of the well-known excess observed by CDF at 
high-$E_T$. In the future (run II) the increase in statistics
will improve the bounds in the  ($\kappa, M_{Z'}$) plane by 
enlarging the upper part of the CDF and D0 areas (or will lead to 
a discovery). For relatively low $M_{Z'}$ values, the main problem
comes from the large systematic errors for "low" $E_T$ jets.
Due to these systematics, at Tevatron, even with a high statistics
it will be difficult
to probe the low $\kappa$ region for $M_{Z'} \leq 400$ GeV
or to close the windows around $M_{Z'} \simeq 300$ and 100 GeV. 
In this respect, as can be seen
from Fig.1, the RHIC-Spin measurements at high luminosity should
allow to cover this region and to get definite conclusions, if 
the new interaction violates parity.

\begin{figure}[t]
\centerline{\psfig{file={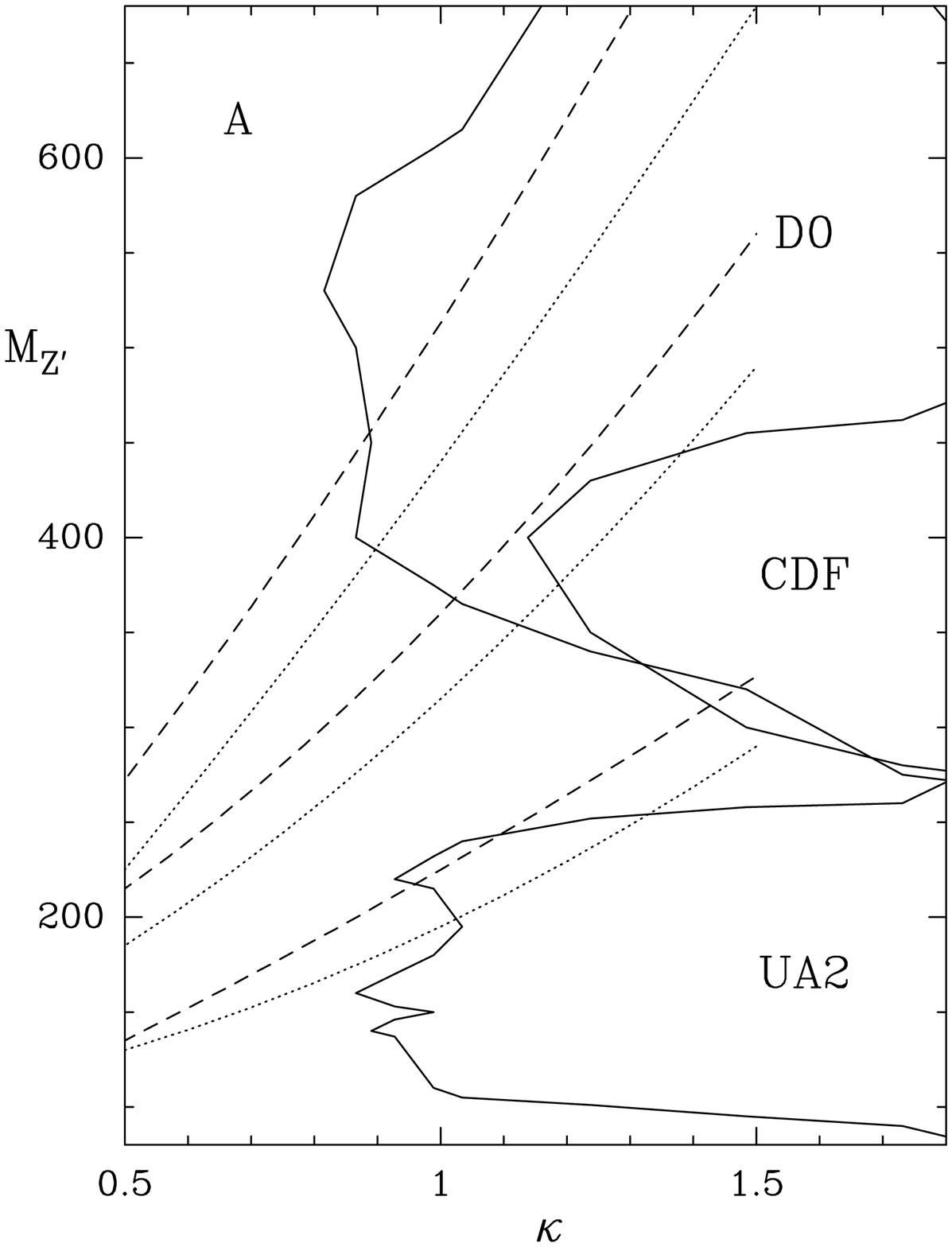},width=10truecm,height=16truecm}}
\caption{Bounds on the parameter space for leptophobic 
flipped SU(5) $Z'$ models (see text).}
    \label{fig1} 
\end{figure}


\section{Comments}
\indent

Concerning experimental uncertainties, with a good knowledge of 
the beam polarization ($\pm 5\%$) and a very good relative luminosity 
measurement ($10^{-4}$), the systematic scale of uncertainty for
a single spin measurement should be of the order of 5\% \cite{Bunce}. 
Hence we have been
conservative in taking $\delta_{syst} = 10\%$.
For instance, one should get  higher limits 
with the former figure : $\Lambda = 9.0 (10.35)$ TeV
with 100 $fb^{-1}$ at 500 (650) GeV, with $\Delta \eta = 1$. The consequences
of a smaller systematical error are more sizeable at high luminosity where the
statistical error becomes very small.

\m

On the theoretical side, the current prejudice is that 
spin asymmetries are much 
less affected than simple cross sections by higher order corrections.
Indeed, recent calculations confirm this simple behaviour.

Concerning SM PV effects, their precise knowledge is mandatory
to extract any signal of New Physics. It has been stressed
in Ref.\cite{BNL2000} that corrections to the QCD-Electroweak interfence
terms, at the order $\alpha_s^2\alpha_W$, might be important in 
the quark-quark channel and also that there were some new contributions
from this order in quark-gluon scattering.
\\
Recently, the authors of Ref.\cite{EMR} have carried out
the calculation of the one-loop weak corrections
to polarized $q-g$ scattering and the corresponding crossed channels.
It appears that the PV effects involving gluons are relatively small, 
i.e. at most 10\% of the tree-level contribution. Moreover, any effect
at the partonic level will not be enhanced by a possibly large
polarization of the gluons, $\Delta G$, because in the large $x$
region which is of interest here the gluon distributions are small.
We have implemented the NLO amplitudes of Ref.\cite{EMR} in our code, 
and we have verified that the corrections on $A_L$ are of the order of 
5\% (7\%) on the whole $E_T$ spectrum at a c.m. energy
of 500 GeV (650 GeV).
It was also straightforward to add the effect of the presence of a 
new \ZP in the one-loop amplitude : it turns out that the contribution
is negligeable.

Concerning $q-q$ scattering, the NLO calculations are not available
but we hope to have them in a not too distant future
\cite{Ross}. However, as shown recently by Vogelsang \cite{BNL2001},
a relatively good estimate of the size of these corrections can be obtained
by performing some gluon resummations. Results of a calculation on $A_L$
at RHIC, after resummation at the Leading-Log level, indicate
a relatively small correction, of the order of 10\% at high $E_T$.
However more precise calculations at the Next-to-Leading-Log level
are necessary to get a definite conclusion.


\section{Conclusions}
Qualitatively new measurements will be allowed by the RHIC-Spin
experiment.
Parity violation searches for physics beyond the 
Standard Model will be competitive with unpolarized searches 
at the Fermilab Tevatron, in particular in the upgraded version
of the machine and of the detector(s). It is worth stressing that
an increase in luminosity of the RHIC $\vec p \vec p$ machine and/or an 
improvement of the angular
coverage of the detectors seem more efficient than an increase in 
energy above $\sqrt s = 600$ GeV.

From now the precise amount of systematic uncertainties is not
accurately known. However experts at RHIC are confident in 
the capacities of polarimetry and luminosity calibrations.
On the other hand, some recent theoretical results indicate that the
tree-level prediction for the SM parity-violating asymmetry
is quite stable. Hence definite results could be obtained from
the measurement of $A_L$ : in particular it has to be emphasized
that the existence of a new weak force between quarks only is not
in contradiction with present data. It might also explain
the small 
discrepancies which still exists between leptonic and hadronic
observables in LEP and SLC results.

Concerning an other possible step for the program, the possibility of 
colliding polarized protons against polarized (or unpolarized) $^3He$ nuclei
has been discussed. This could allow to measure some spin asymmetries
in $\vec p$-$n$ and/or $\vec p$-$\vec n$ collisions and also possibly
in $\vec n$-$\vec n$ collisions via polarized $^3He$-$^3He$ collisions. 
In this case a new charged vector boson (e.g. a massive right-handed $W_R$) could also mediate some visible effects (see  Ref.\cite{TVWprime}).

\m
\no {\bf Acknowledgments}\\ 
J.M.V. acknowledges the warm hospitality at the RIKEN-BNL Research center where
part of this work has been performed. Thanks are due to 
G. Bunce, G. Eppley, S. Moretti, D.A. Ross, N. Saito, J. Soffer and W. Vogelsang 
for fruitful discussions.






\end{document}